\PassOptionsToPackage{table}{xcolor}
\documentclass[sigconf]{acmart}

\usepackage{amsmath}
\usepackage{amsfonts}
\usepackage{booktabs}
\usepackage{array}
\usepackage{multirow}
\usepackage{graphicx}
\usepackage{pifont}
\usepackage{placeins}  
\usepackage{tabularx}
\usepackage{subcaption}
\usepackage{enumitem}

\newcommand{\full}{\(\bullet\)}      
\newcommand{\none}{\(\circ\)}        
\newcommand{\parti}{\(\odot\)}       

\makeatletter
\renewcommand{\@affiliationfont}{\normalsize\normalfont}
\makeatother
\renewcommand\footnotetextcopyrightpermission[1]{}
\setcopyright{none}

\copyrightyear{2026}
\acmYear{2026}
\acmDOI{}
\acmISBN{}

\acmConference[ICCAD '26]{IEEE/ACM International Conference on Computer-Aided Design}{November 8--12, 2026}{San Jose, CA, USA}
\acmBooktitle{Proceedings of the IEEE/ACM International Conference on Computer-Aided Design (ICCAD '26), November 8--12, 2026, San Jose, CA, USA}

\title{StateTune: Transforming LLM-Assisted EDA Flow Tuning into a Stateful, Closed-Loop Process}

\author{KunLong Li}
\authornote{Both authors contributed equally to this research.}
\affiliation{%
  \institution{Fudan University}
  \city{Shanghai}
  \country{China}
}
\email{klli24@m.fudan.edu.cn}

\author{Shangshang Yao}
\authornotemark[1]
\affiliation{%
  \institution{Independent Researcher}
  \city{Shanghai}
  \country{China}
}
\email{yaoshangshang96@outlook.com}

\author{Su Zheng}
\affiliation{%
  \institution{\mbox{Chinese University of Hong Kong}}
  \city{Hong Kong}
  \country{China}
}
\email{szheng22@cse.cuhk.edu.hk}

\author{Lingli Wang}
\authornote{Corresponding author.}
\affiliation{%
  \institution{Fudan University}
  \city{Shanghai}
  \country{China}
}
\email{llwang@fudan.edu.cn}

\begin{document}

\begin{abstract}
EDA flow parameter tuning is critical for quality-of-results~(QoR), yet the parameter space is large, tightly coupled, and full evaluations are prohibitively expensive.
Prior LLM-assisted tuners mainly use the LLM as an external proposer with transient working context; we instead present \textbf{StateTune}, which reformulates LLM-assisted EDA tuning as a closed-loop, state-carrying process.
Its optimizer state is a typed, evidence-gated \emph{persistent optimization memory} that is updated by every evaluation and shared between candidate generation and budget allocation. On top of this optimizer state, an expected hypervolume improvement (EHVI)-guided, runtime-aware promotion policy ranks quick-stage candidates by expected Pareto frontier gain per unit of runtime cost.
Evaluated on a Cadence industrial flow across six benchmark blocks (two technology nodes \(\times\) three designs), against five baselines including LLM+retrieval-augmented generation (RAG) and preference-based Bayesian optimization (BO) tuners, StateTune achieves the strongest final hypervolume on all six benchmark blocks, showing a stable improvement in frontier quality across the full matrix; it also matches or surpasses the strongest baselines on worst negative slack (WNS), area, and power across the same set.
Ablation shows persistent memory is the largest contributor: removing it costs 58.5\% of the hypervolume.
Dedicated analyses of evidence-gating sensitivity, memory poisoning, cross-design transfer, and three-seed reproducibility (CV\,\(<\)\,7\% on five of six blocks) further validate the memory design.
\end{abstract}

\keywords{EDA flow tuning, large language models, persistent memory, multi-fidelity optimization, EHVI, physical design}

\maketitle

\section{Introduction}

Modern physical-implementation flows expose many interacting knobs across floorplanning, placement, clock-tree synthesis (CTS), and routing.
These knobs are tightly coupled: even modest changes to utilization, timing effort, or density can materially alter timing, area, and power---or cause a run to fail.
Because full register-transfer-level (RTL)-to-route evaluations take hours, a practitioner's tuning budget is strictly limited.
This makes EDA flow tuning a \emph{budgeted} decision problem that must simultaneously address three challenges: proposing good configurations, learning from each evaluation, and deciding which candidates deserve the scarce full-flow budget.

Automated methods have made significant progress.
METRICS2.1~\cite{jung2021metrics21} standardized no-human-in-the-loop evaluation, and algorithmic advances such as PTPT~\cite{geng2023ptpt}, REMOTune~\cite{zheng2023remotune}, RankTuner~\cite{xu2024ranktuner}, and FlowTuner~\cite{liang2021flowtuner} have demonstrated multi-objective Bayesian optimization, trust-region search, preference-based tuning, and cross-stage knowledge transfer.
However, these methods treat the EDA tool as a black box: the design knowledge uncovered during tuning---which parameter ranges trigger failures, which interactions help, and why configurations succeed or fail---remains implicit in the optimizer state and is discarded after each run, so subsequent iterations and new runs cannot benefit from earlier insights.

Large language models (LLMs) have been adopted in EDA for tool interaction, documentation retrieval, and design-space exploration~\cite{wu2024chateda,he2025futuremirage,zhong2024llm4eda}.
For flow tuning specifically, CROP combined circuit-level retrieval-augmented generation (RAG) with LLM-guided search~\cite{pan2025crop} and ORFS-agent demonstrated iterative LLM-based parameter optimization for OpenROAD~\cite{ghose2025orfsagent}, showing that LLMs can leverage domain knowledge to generate stronger proposals than blind numerical search.
However, these systems primarily rely on transient search context assembled at inference time rather than a structured, evolving memory that distills experience into reusable artifacts.
Moreover, they focus exclusively on candidate \emph{generation} and leave the complementary \emph{promotion} decision---which of several proposals deserves the scarce full-flow budget---to ad-hoc filtering or unselective evaluation.

StateTune takes a different route. Rather than treating the LLM as a smarter proposer bolted onto a standard tuning loop, we \emph{reformulate} LLM-assisted EDA tuning as a closed-loop, state-carrying process whose optimizer state is an explicit, typed, evidence-gated \textbf{persistent optimization memory} updated by every evaluation and shared between candidate generation and budget allocation (Section~\ref{sec:memory}).
On top of this state, an \textbf{expected hypervolume improvement (EHVI)-guided, runtime-aware promotion} policy ranks candidates by expected Pareto frontier gain per unit of runtime cost, directing the expensive budget toward the most promising proposals.
We evaluate StateTune on a Cadence Genus/Innovus industrial flow across ASAP7 and NanGate45 with benchmark blocks covering JPEG, AES, and IBEX, and compare against five baselines including two strong LLM/retrieval-based tuners.
Our code is open-sourced in \href{https://github.com/C-YuLong/stateTune}{our repository}.

The main contributions are:
\begin{itemize}[leftmargin=0pt]
    \item \textbf{A closed-loop reformulation of LLM-assisted EDA tuning} in which the optimizer state is a typed, evidence-gated persistent memory \(\mathcal{M}_t\) that jointly drives candidate generation and budget allocation. Evidence gating prevents knowledge poisoning~\cite{chen2024agentpoison}; ablation shows this is the largest single contributor (HV \(-58.5\%\) when removed) (Section~\ref{sec:memory}).
    \item \textbf{EHVI-guided, runtime-aware promotion} that ranks quick-stage candidates by expected Pareto frontier gain per unit of runtime risk, directly reading from \(\mathcal{M}_t\) so that improvements to memory propagate to budget allocation through a shared state object.
    \item \textbf{Empirical validation on a \(2{\times}3\) benchmark matrix} (two technology nodes \(\times\) three designs), with a six-way comparison against BO\,(qEHVI), Optuna-TPE, Random, RankTuner~\cite{xu2024ranktuner}, and CROP~\cite{pan2025crop}. StateTune attains the best WNS, best power, and best hypervolume on all six benchmark blocks, together with the best or tied-best area on all six. Component-level evidence from five ablation variants, dedicated analyses of evidence-gating sensitivity and memory poisoning, and a three-seed reproducibility study support these results.
\end{itemize}

\begin{figure*}[ht]
    \centering
    \includegraphics[width=0.88\textwidth]{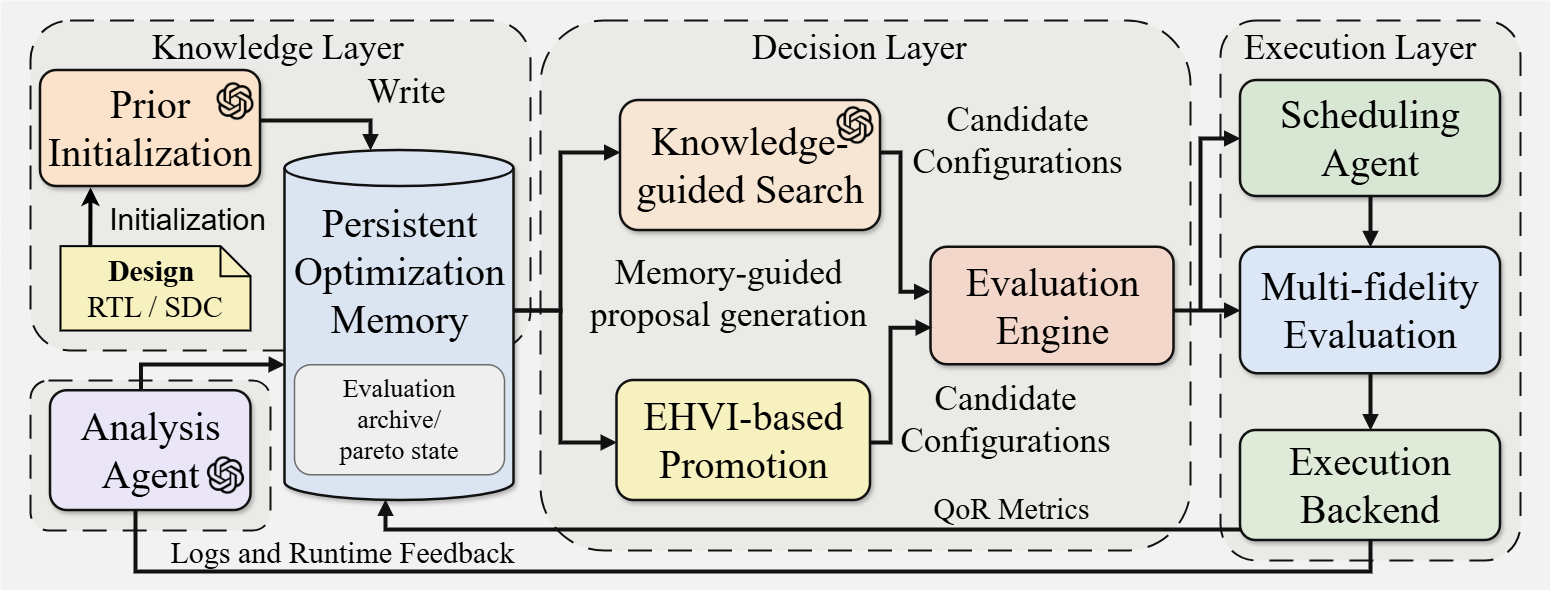}
    \caption{Overall workflow of StateTune. The knowledge layer (left) builds and updates persistent optimization memory from design inputs and evaluation feedback via Prior Initialization and an Analysis Agent. The decision layer (center) generates candidate configurations through Knowledge-guided Search and ranks them via EHVI-based Promotion; an Evaluation Engine arbitrates the final selection. The execution layer (right) schedules multi-fidelity evaluations through a Scheduling Agent and returns QoR metrics. The persistent optimization memory connects all three layers.}
    \label{fig:overview}
\end{figure*}


\section{Related Work}

We review three lines of work that StateTune builds upon: multi-objective optimization under budget constraints, structured memory for LLM agents, and automated EDA flow tuning.

\subsection{Multi-Objective BO and Cost-Aware Search}

Bayesian optimization (BO) models objectives via Gaussian process (GP) surrogates and uses acquisition functions to guide search~\cite{shahriari2016bo}.
In multi-objective settings, qEHVI~\cite{daulton2020qehvi} quantifies the expected volume gained by adding a new point to the Pareto frontier, providing a principled criterion for improving the entire frontier rather than a single metric.
When evaluations are expensive, multi-fidelity strategies can improve sample efficiency by exploiting cheaper proxy evaluations before committing to full runs; FABOLAS and BOHB demonstrated this for hyperparameter optimization~\cite{klein2017fabolas,falkner2018bohb}.
Optuna applies tree-structured Parzen estimators (TPE) to multi-objective problems~\cite{akiba2019optuna,bergstra2011algorithms}.
StateTune adopts this budget-aware perspective but embeds it inside an EDA-specific loop where quick-stage memory, accumulated design knowledge, and runtime-aware promotion interact tightly.

\subsection{Structured Memory for LLM Agents}

A growing body of work shows that LLM agents benefit from structured memory beyond raw conversation history.
Reflexion~\cite{shinn2023reflexion} introduced verbal reinforcement learning via trajectory-level reflection; A-MEM~\cite{xu2025amem} organizes knowledge notes via Zettelkasten-style indexing; MemoryOS~\cite{kang2025memoryos} designs a hierarchical short/mid/long-term memory; and CFGM~\cite{yang2025cfgm} grounds memories at multiple granularities for planning.
RSR~\cite{zhao2025rsr} decomposes agent memory into retrieval, scheduling, and reflection stages, demonstrating that episodic memory with a reflection loop improves task performance in sequential decision-making.

A key concern with persistent memory is \emph{quality}: incorrect or outdated information can degrade later decisions.
AgentPoison~\cite{chen2024agentpoison} formalizes this risk in the security context, and in optimization settings the same risk arises self-inflicted when early-stage LLM diagnoses based on insufficient data produce incorrect rules.

StateTune differs from these general-purpose memory frameworks in two structural respects.
First, its memory is \emph{typed and evidence-gated}: each artifact (hard rule, soft heuristic, sensitivity pattern, failure summary) has an explicit type and activation condition, with hard rules requiring \(k{=}12\) corroborating full-valid observations before activation; the memory-poisoning experiment in Section~\ref{sec:poisoning} confirms that removing this gating causes incorrect rules to accumulate and degrades HV by 43.4\%.
Second, the memory is \emph{shared} between candidate generation and a downstream budget-allocation policy, so that an improvement to a rule or sensitivity pattern propagates to both proposal quality and promotion ranking through a single state object---a coupling that RSR's episodic reflection and A-MEM's note indexing do not provide (Table~\ref{tab:capability}).

\subsection{EDA Flow Tuning and LLM-Assisted Search}

Automated EDA flow tuning has evolved from reusable infrastructure to specialized algorithmic search.
METRICS2.1~\cite{jung2021metrics21} established a no-human-in-the-loop evaluation stack; PTPT~\cite{geng2023ptpt} formulated tuning as multi-task GP-based BO; REMOTune~\cite{zheng2023remotune} scaled to high dimensions via random embedding; RankTuner~\cite{xu2024ranktuner} introduced preference-based BO for noisy QoR; FlowTuner~\cite{liang2021flowtuner} transferred knowledge across stages.
These methods advance numerical search but do not externalize the non-numerical knowledge---failure modes, parameter interactions, design heuristics---that emerges during tuning.

On the LLM side, ChatEDA~\cite{wu2024chateda} demonstrated autonomous EDA tool interaction; CROP~\cite{pan2025crop} combined circuit-level RAG with LLM-guided tuning; ORFS-agent~\cite{ghose2025orfsagent} showed iterative LLM-based parameter optimization for OpenROAD.
AstroTune~\cite{wang2026astrotune} integrated AST-based retrieval with stage-wise tournament BO, combining retrieval-augmented LLM proposals with a BO backbone; however, its retrieval feeds only the proposal stage and does not parameterize a downstream budget-allocation policy.
Recent surveys provide broader coverage~\cite{he2025futuremirage,zhong2024llm4eda}.
These systems keep search context largely transient, and none externalizes a \emph{typed, evidence-gated} optimizer state that is shared between candidate generation and budget allocation---the three structural gaps (typed state, evidence gating, shared promotion) that StateTune fills through a single persistent memory object \(\mathcal{M}_t\).
\begin{table}[!t]
\centering
\caption{Capability comparison of EDA flow tuning methods.
\full\,=\,supported; \none\,=\,not supported; \parti\,=\,partial.}
\label{tab:capability}
\small
\renewcommand{\arraystretch}{1.12}
\setlength{\tabcolsep}{2.6pt}

\begin{tabular}{p{1.78cm} c c c c c c c}
\toprule
Method
& \shortstack{LLM\\search}
& RAG
& \shortstack{Memory\\type}
& Gate
& \shortstack{Multi-\\fid.}
& \shortstack{Multi-\\obj.}
& \shortstack{Cross-\\design} \\
\midrule
PTPT~\cite{geng2023ptpt}
  & \none & \none & -- & \none & \none & \full & \none \\
FlowTuner~\cite{liang2021flowtuner}
  & \none & \none & -- & \none & \full\(^a\) & \none & \none \\
RankTuner~\cite{xu2024ranktuner}
  & \none & \none & -- & \none & \none & \full & \none \\
CROP~\cite{pan2025crop}
  & \full & \full\(^b\) & \shortstack{Prompt\\hist.} & \none & \none & \none & \none \\
ORFS-agent~\cite{ghose2025orfsagent}
  & \full & \none & \shortstack{Prompt\\hist.} & \none & \none & \full\(^c\) & \none \\
AstroTune~\cite{wang2026astrotune}
  & \parti\(^d\) & \full & -- & \none & \full & \full & \none \\
RSR~\cite{zhao2025rsr}
  & \full & \parti\(^f\) & \shortstack{Episodic} & \none & \none & \none & \none \\
\midrule
\textbf{StateTune (Ours)}
  & \textbf{\full} & \textbf{\full\(^e\)} & \textbf{Persistent}
  & \textbf{\full} & \textbf{\full} & \textbf{\full} & \textbf{\full} \\
\bottomrule
\end{tabular}

\vspace{3pt}
\footnotesize
\(^{a}\) Stage transfer via jump-start/early-stop.\
\(^{b}\) Design-level retrieval from similar circuits.\
\(^{c}\) Weighted-sum / constrained multi-objective prompts.\
\(^{d}\) Retrieval augmented by LLM; BO handles search.\
\(^{e}\) Design- and run-level retrieval via RAG-EDA~\cite{pu2024rageda}.\
\(^{f}\) Retrieval of past reflection traces; no EDA-specific RAG.
\end{table}

\section{Method}

This section describes the three layers of StateTune---knowledge, decision, and execution---and details the persistent optimization memory and EHVI-guided promotion that connect them.

\subsection{Problem Setting and Overview}

We consider a budgeted multi-objective tuning problem over a parameter vector \(x \in \mathcal{X}\).
A full evaluation \(e(x)\) returns a QoR vector \(\mathbf{f}(x) = \bigl(f_{\text{WNS}}(x),\, f_{\text{area}}(x),\, f_{\text{power}}(x)\bigr)\), where WNS denotes worst negative slack, runtime \(c(x)\), and terminal status \(\delta(x) \in \{0,1\}\).
The tuning goal is to approximate the Pareto frontier
\begin{equation}
\mathcal{P}^{*} = \bigl\{x \in \mathcal{X} : \nexists\, x'\text{ s.t.\ } \mathbf{f}(x') \succ \mathbf{f}(x)\bigr\}
\end{equation}
so as to maximize the dominated hypervolume \(\mathrm{HV}(\mathcal{P},\,\mathbf{r})\) with respect to a fixed reference point \(\mathbf{r}\), subject to a total runtime budget \(\sum_i c(x_i) \le B\).
Because full evaluations are expensive, the system first executes a cheaper quick stage (cost \(c_q \ll c_f\)) and promotes only selected candidates to the full stage.
The quick-stage cutoff point is configurable; in our experiments we use CTS (clock-tree synthesis) as the quick-stage endpoint, which provides richer timing signal than placement alone (Section~\ref{sec:setup}).
StateTune tunes 19 parameters spanning floorplanning, placement, pre-CTS/CTS, and routing (Table~\ref{tab:params}).
The optimizer state is a typed, evidence-gated persistent optimization memory \(\mathcal{M}_t\) carried across iterations, rather than a growing prompt transcript (detailed in Section~\ref{sec:memory}).

StateTune is organized into three interacting layers (Figure~\ref{fig:overview}).

The \textbf{knowledge layer} manages the persistent optimization memory.
A Prior Initialization module initializes a design prior from Synopsys Design Constraints (SDC), RTL characteristics, and RAG-retrieved tool documentation.
As evaluations complete, an Analysis Agent periodically runs failure pattern analysis, parameter sensitivity analysis (including threshold-effect detection), and prior refinement, feeding curated results back into memory.

The \textbf{decision layer} uses the memory to propose and filter candidate configurations.
A Knowledge-guided Search module queries the memory and selects among five search modes---\emph{explore}, \emph{exploit}, \emph{diversify}, \emph{repair}, and \emph{promote}---based on recent progress, frontier diversity, and failure patterns (detailed in Section~\ref{sec:search}).
Chain-of-thought (CoT) reasoning distills the rich context into structured diagnoses, and a non-reasoning model formats final parameter vectors.
An EHVI-based Promotion module then ranks the generated candidates by expected Pareto frontier gain per unit of runtime cost (Section~\ref{sec:promotion}), and an Evaluation Engine arbitrates the final selection for full evaluation.

The \textbf{execution layer} manages multi-fidelity scheduling through a Scheduling Agent.
Quick-stage evaluations---CTS in our experiments, though the cutoff is configurable---are cheap and run continuously.
Full-stage evaluations through routing are expensive and allocated via EHVI-guided promotion.
Results and runtime measurements flow back to the knowledge layer, closing the loop.

The central design choice is that \(\mathcal{M}_t\) is persistent and reusable: knowledge from early iterations directly shapes later proposals and promotion decisions.
Three properties distinguish \(\mathcal{M}_t\) from prior optimizer states: it stores typed artifacts (rules, sensitivities, failure patterns) that a GP posterior or transient context cannot represent robustly; hard rules require \(k{=}12\) corroborating full-valid observations before activation; and the same shared object parameterizes both candidate generation and EHVI-guided promotion.
Ablation confirms that this explicit optimizer state is the largest single contributor to final hypervolume (Section~\ref{sec:ablation}).

\subsection{Persistent Optimization Memory}
\label{sec:memory}

\begin{figure}[ht]
    \centering
    \includegraphics[width=\linewidth]{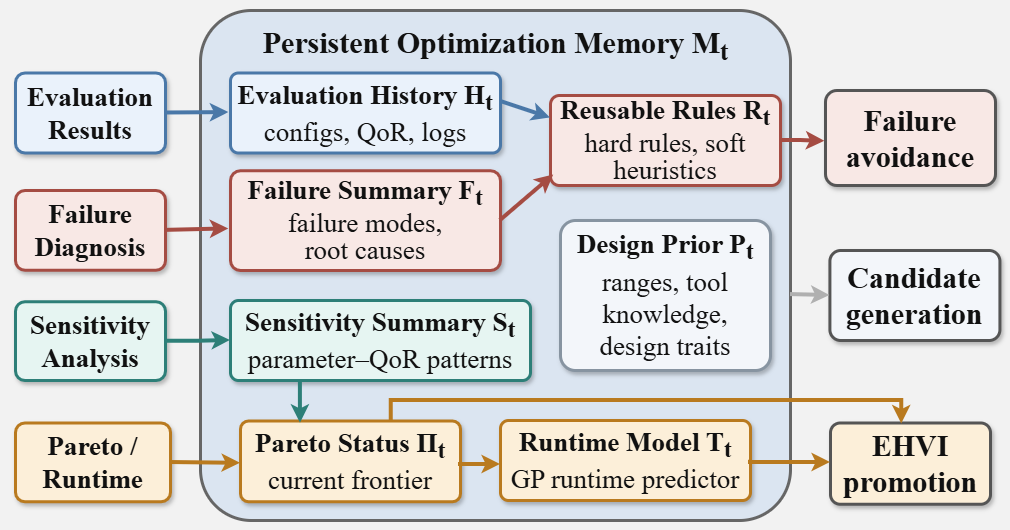}
    \caption{Structure of the persistent optimization memory \(\mathcal{M}_t\). Evaluation results, failure diagnoses, and sensitivity analyses feed into six components. The design prior \(P_t\) conditions generation; reusable rules \(R_t\) guide failure avoidance; Pareto status \(\Pi_t\) and the runtime model \(T_t\) drive EHVI-based promotion.}
    \label{fig:mem}
\end{figure}

The persistent optimization memory at iteration \(t\) is
\begin{equation}
\mathcal{M}_t = \{P_t, R_t, F_t, S_t, H_t, \Pi_t, T_t\},
\end{equation}
where \(P_t\) is the design prior, \(R_t\) contains reusable rules, \(F_t\) is a failure summary, \(S_t\) is a sensitivity summary, \(H_t\) is the evaluation history, \(\Pi_t\) is Pareto status, and \(T_t\) is the runtime model.
Table~\ref{tab:memory} describes each component.

\begin{table}[t]
\centering
\caption{Persistent memory components.}
\label{tab:memory}
\scriptsize
\begin{tabular}{p{0.18\linewidth} p{0.42\linewidth} p{0.28\linewidth}}
\toprule
Component & Content & Role \\
\midrule
Prior \(P_t\) & Parameter ranges, tool knowledge, design characteristics & Conditions proposals \\
Rules \(R_t\) & Evidence-gated hard rules \& soft heuristics & Filters proposals \\
Failure \(F_t\) & Categorized failures with root-cause analysis & Guides repair \\
Sensitivity \(S_t\) & Parameter--QoR correlations \& threshold effects & Prioritizes knobs \\
History \(H_t\) & All configs, QoR, terminal status & LLM context \& GP fit \\
Pareto \(\Pi_t\) & Current non-dominated frontier & EHVI targets \\
Runtime \(T_t\) & \(k\)NN runtime predictions & Promotion cost \\
\bottomrule
\end{tabular}
\end{table}

The prior \(P_t\) is initialized from SDC specifications, RTL characteristics, and RAG-retrieved tool documentation, then periodically refined as evaluations accumulate so that it adapts to each design's characteristics.
The failure summary \(F_t\) periodically categorizes failed runs by type and associated parameter combinations, injecting dominant patterns as compact text into the search prompt in \emph{repair} mode.

\textbf{Sensitivity analysis.}
The sensitivity summary \(S_t\) combines Spearman rank correlations (with a weighted impact score across WNS, area, power) and a threshold-effect detector that flags nonlinear cliff effects invisible to correlation analysis alone.
Both outputs are formatted as compact text and injected into the proposal prompt; concrete examples appear in Table~\ref{tab:rules}.

\textbf{Evidence gating and memory quality control.}
LLM-emitted rules based on insufficient data can be incorrect; if stored directly, such errors accumulate and progressively degrade search quality---a self-inflicted variant of the memory-poisoning risk documented in LLM agent systems~\cite{chen2024agentpoison}.
Evidence gating prevents this: hard rules require at least \(k{=}12\) corroborating full-valid evaluations before activation (Section~\ref{sec:ksensitivity} and Figure~\ref{fig:k_ablation} validate this threshold), the number of active hard rules is capped at 8, and a periodic trim removes stale or contradicted entries.
Soft heuristics act as overridable directional preferences.
The memory-poisoning experiment in Section~\ref{sec:poisoning} confirms that removing gating degrades HV by 43.4\%.

\begin{table}[t]
\centering
\caption{Knowledge artifacts automatically generated by StateTune.}
\label{tab:rules}
\scriptsize
\begin{tabular}{p{0.13\linewidth} p{0.55\linewidth} p{0.20\linewidth}}
\toprule
Type & Content & Evidence \\
\midrule
Hard rule & Ban \texttt{CORE\_UTIL=0.78} with \texttt{CORE\_MARGIN=0}: catastrophic WNS (\(-32\)k) from placement congestion. & Multiple failures \\
Soft heur. & Increase \texttt{CORE\_UTIL} \& \texttt{PLACE\_MAX\_DENSITY}; keep \texttt{CORE\_MARGIN} low. WNS corr.\ +0.24. & Sensitivity \\
Sensitivity & \texttt{CTS\_POST\_OPT} corr.\ \(-\)0.19 with WNS; its heuristics degrade setup timing. & 365 runs \\
Prior upd. & \texttt{ASPECT\_RATIO}: 1.2\(\to\)1.0 (square). Rectangular shapes worse for bus-heavy datapath. & 12 full-valid obs. \\
\bottomrule
\end{tabular}
\end{table}

\subsection{Knowledge-Guided Search}
\label{sec:search}

The knowledge-guided search module uses the persistent memory to generate context-aware candidate configurations.
For each iteration, it selects among five search modes based on recent progress, frontier diversity, and failure patterns:
\emph{explore} broadens coverage of the parameter space;
\emph{exploit} refines configurations near the current best;
\emph{diversify} increases Pareto frontier spread;
\emph{repair} targets known failure modes using the failure summary \(F_t\) with corrective adjustments;
and \emph{promote} nominates strong quick-stage candidates for full evaluation.
Mode selection follows a heuristic priority: \emph{repair} is triggered when the recent failure rate exceeds a threshold; \emph{promote} is selected when the promotion pool contains sufficiently strong candidates; among the remaining modes, the system alternates based on frontier stagnation (favoring \emph{diversify}) and convergence signals (favoring \emph{exploit}).
The proposal prompt is conditioned on the parameter space, recent history, Pareto points, the updated design prior, failure and sensitivity summaries, and stage-specific constraints.

\textbf{Dual-model inference.}
StateTune uses CoT to organize RAG-retrieved documentation, evaluation histories, and parameter interactions into structured failure and sensitivity diagnoses, while a non-reasoning model formats parsable candidate vectors.
We use a strong open-weight reasoning model and a fast generation model (Table~\ref{tab:setup}) to balance analytical depth with output reliability.
Ablation shows that removing CoT reduces HV by 31.5\% (Section~\ref{sec:ablation}).
LLM calls are made only when the candidate buffer is depleted, not at every iteration, keeping token consumption bounded.

\textbf{Candidate buffer and frontier-driven refresh.}
Each LLM call produces multiple candidate configurations that are stored in a buffer and consumed one-by-one in subsequent iterations.
To ensure that proposals reflect the latest optimization state, the buffer is \emph{flushed} whenever a new Pareto-optimal point is discovered or the hypervolume improves: all remaining buffered candidates are discarded, forcing the next iteration to issue a fresh LLM call conditioned on the updated history and frontier.
This mechanism prevents the system from wasting iterations on candidates generated from stale context.
\textbf{Stage-aware conditioning.}
Quick-stage proposals are biased toward placement-effective parameters, since routing knobs have little visible effect at the quick stage.

\textbf{Rule emission and statistical validation.}
The proposal module emits hard rules and soft heuristics stored in memory.
Representative artifacts include banning parameter combinations repeatedly associated with catastrophic congestion, promoting sensitivity-backed density adjustments, and updating design priors when repeated full-valid observations support a tighter setting.
Evidence gating (Section~\ref{sec:memory}) prevents premature knowledge from entering memory.
Specifically, each candidate rule is validated \emph{post hoc} against accumulated full-valid observations: the system partitions the observation set into rule-conforming and rule-violating groups and applies a one-sided Mann--Whitney \(U\) test (\(\alpha = 0.05\)) to determine whether the conforming group exhibits statistically significantly better QoR.
Only rules that pass this test and have been corroborated by at least \(k\) observations are promoted to active status.
Across the ASAP7 JPEG run, only 1 of 13 hard rules and 30 of 101 soft heuristics were validated; 6 soft heuristics were contradicted.
This noise is precisely why evidence gating is necessary.

\subsection{EHVI-Guided, Runtime-Aware Promotion}
\label{sec:promotion}

LLM-based search generates candidates of varying quality, so evaluating them indiscriminately wastes scarce full-evaluation budget.
StateTune therefore ranks candidates by expected Pareto frontier gain per unit of runtime cost and promotes only the most promising ones.

To leverage the abundant quick-stage observations alongside scarce full-stage data, StateTune fits a multi-fidelity GP surrogate for each QoR objective.
Quick-stage and full-stage histories are jointly modeled using a binary fidelity indicator (\(z{=}0\) for quick, \(z{=}1\) for full); when insufficient full-stage data makes the multi-fidelity fit unreliable, the system falls back to a standard single-fidelity GP trained on full-stage data only.
At scoring time, candidates are projected to the full-fidelity surface for each QoR objective \(f_k(x)\), \(k = 1, \dots, K\):
\begin{equation}
f_k(x) \sim \mathcal{GP}\bigl(\mu_k(x),\, \sigma_k^2(x)\bigr),
\end{equation}
where \(\mu_k(x)\) and \(\sigma_k^2(x)\) are the posterior mean and variance given \(H_t\), and \(K=3\) (WNS, area, power).

Given the current Pareto frontier \(\Pi_t\) and reference point \(\mathbf{r}\), EHVI quantifies expected gain:
\begin{equation}
\mathrm{EHVI}(x) =
\mathbb{E}_{\mathbf{f}(x)}
\bigl[
\mathrm{HV}(\Pi_t \cup \{\mathbf{f}(x)\},\, \mathbf{r})
-
\mathrm{HV}(\Pi_t,\, \mathbf{r})
\bigr],
\end{equation}
following the qEHVI formulation~\cite{daulton2020qehvi}.

\textbf{Runtime cost model.}
Predicting full-stage runtime is challenging because few full evaluations are available early on.
GP-based runtime models require careful kernel selection and sufficient data for reliable posteriors.
StateTune uses a \(k\)-nearest-neighbor (\(k\)NN) estimator over \(H_t\) for runtime prediction, which is stable with small sample sizes, requires no hyperparameter tuning, and provides a natural uncertainty estimate via neighbor variance.
Since promotion only needs to \emph{rank} candidates correctly rather than predict absolute runtimes, \(k\)NN's simplicity is well matched to the task.
The \(k\)NN model estimates runtime as:
\begin{equation}
\hat{t}(x) = \frac{1}{k}\sum_{i=1}^{k} t(x_i),
\end{equation}
where \(x_i\) are the \(k\) nearest evaluated configurations.
An uncertainty-aware upper confidence bound is:
\begin{equation}
\hat{t}_{\mathrm{ucb}}(x) = \hat{t}(x) + \beta \cdot s_t(x),
\end{equation}
where \(s_t(x)\) is the sample standard deviation and \(\beta > 0\) controls cost pessimism.
Promotion is ranked by:
\begin{equation}
\mathrm{score}(x) = \frac{\mathrm{EHVI}(x)}{\hat{t}_{\mathrm{ucb}}(x) + \epsilon},
\end{equation}
where \(\epsilon > 0\) is a stabilizing constant.
This score optimizes expected frontier gain per unit of runtime risk, directing the budget toward cost-effective frontier improvements.

Empirically, the EHVI-guided promotion concentrates full-evaluation budget on strong quick-stage candidates: across the reported benchmark set, the fraction of promoted candidates that rank in the historical top-20 of quick-stage evaluations averages \(0.96 \pm 0.08\) (\(p < 0.001\) vs.\ the 0.5 null, paired \(t\)-test), confirming that promotion functions as a quality filter.
Ablation confirms this mechanism's importance: removing it reduces the top-region hit rate from 34.29\% to 8.57\% (Section~\ref{sec:ablation}).

\section{Experimental Setup}
\label{sec:setup}

This section describes the benchmarks, baselines, and evaluation protocol.

\begin{table}[t]
\centering
\caption{Experimental setup and LLM overhead.}
\label{tab:setup}
\scriptsize
\begin{tabular}{p{0.22\linewidth} p{0.68\linewidth}}
\toprule
\multicolumn{2}{l}{\textbf{Benchmarks \& Flow}} \\
\midrule
Flow & Cadence Genus~21.17 / Innovus~21.18 \\
Benchmarks & JPEG / AES / IBEX on both ASAP7~\cite{clark2016asap7} (7\,nm) and NanGate45~\cite{stine2007freepdk} (45\,nm) \\
Objectives & Maximize WNS; minimize area \& power \\
Fidelities & Quick = CTS (configurable); Full = route \\
Parameters & 19 across 4 stages (Table~\ref{tab:params}) \\
\midrule
\multicolumn{2}{l}{\textbf{Budget \& Hardware}} \\
\midrule
Main budget & 192\,thread\(\cdot\)h\(^{\dagger}\) (4 threads \(\times\) 48\,h) \\
Ablation budget & 48\,thread\(\cdot\)h (1 thread \(\times\) 48\,h) \\
CPU & Intel Xeon Plat.\ 8354H @ 3.10\,GHz \\
\midrule
\multicolumn{2}{l}{\textbf{Models \& Baselines}} \\
\midrule
CoT model & Qwen3-235B-A22B-Thinking-2507 \\
Non-reasoning & DeepSeek-V3.2 \\
RAG framework & RAG-EDA~\cite{pu2024rageda} \\
Baselines & BO\,(qEHVI), Optuna-TPE, Random, RankTuner\,(ICCAD'24)~\cite{xu2024ranktuner}, CROP\,(ICCAD'25)~\cite{pan2025crop} \\
Ablations & w/o KA, w/o Persist., w/o EHVI, w/o RAG, w/o CoT \\
\midrule
\multicolumn{2}{l}{\textbf{LLM Overhead}} \\
\midrule
CoT analysis & \(\sim\)15\,s/call, 1 call/iter, \(<\)2\% of wall time \\
RAG + gen. & \(\sim\)5\,s/call, 1--2 calls/iter, \(<\)1\% \\
Quick EDA & \(\sim\)8\,min, \(\sim\)85\% of wall time \\
Full EDA & \(\sim\)45\,min, when promoted \\
Total tokens & \(\sim\)6.0\,M per 6-benchmark suite (74\% prompt, 26\% completion; 81\% DeepSeek-V3.2, 19\% Qwen3-235B) \\
\bottomrule
\end{tabular}

\vspace{2pt}
\footnotesize
\(^{\dagger}\) One thread\(\cdot\)hour denotes one EDA worker thread running for one hour; this measures physical-thread occupancy rather than full-CPU-core time.
\end{table}

\begin{table}[t]
\centering
\caption{Tuned parameters. C=continuous; Cat=categorical; B=binary.}
\label{tab:params}
\scriptsize
\begin{tabular}{p{0.06\linewidth} p{0.42\linewidth} p{0.08\linewidth} p{0.28\linewidth}}
\toprule
Stg & Parameter & Type & Range \\
\midrule
\multirow{3}{*}{FP}
 & \texttt{CORE\_UTIL}              & C & \([0.55,\,0.78]\) \\
 & \texttt{ASPECT\_RATIO}           & C & \([0.80,\,1.20]\) \\
 & \texttt{CORE\_MARGIN}            & Int & \([0,\,16]\) \\
\midrule
\multirow{6}{*}{PL}
 & \texttt{PLACE\_CONG\_EFFORT}     & Cat & low/med/high \\
 & \texttt{PLACE\_TIMING\_EFFORT}   & Cat & med/high \\
 & \texttt{PLACE\_UNIFORM\_DENSITY} & Cat & true/false \\
 & \texttt{PLACE\_IO\_PINS\_AWARE}  & Cat & true/false \\
 & \texttt{TD\_PLACE}               & B & \(\{0,1\}\) \\
 & \texttt{PLACE\_MAX\_DENSITY}     & C & \([0.75,\,0.95]\) \\
\midrule
\multirow{6}{*}{CTS}
 & \texttt{PRECTS\_SETUP\_OPT}      & B & \(\{0,1\}\) \\
 & \texttt{PRECTS\_HOLD\_OPT}       & B & \(\{0,1\}\) \\
 & \texttt{DRV\_EFFORT}             & Cat & low/med/high \\
 & \texttt{RUN\_CTS}                & B & \(\{0,1\}\) \\
 & \texttt{CTS\_TARGET\_SKEW}       & C & \([0.03,\,0.15]\) \\
 & \texttt{CTS\_POST\_OPT}          & B & \(\{0,1\}\) \\
\midrule
\multirow{4}{*}{RT}
 & \texttt{ROUTE\_TIMING\_DRIVEN}   & Cat & true/false \\
 & \texttt{ROUTE\_SI\_DRIVEN}       & Cat & true/false \\
 & \texttt{ROUTE\_EFFORT}           & Cat & low/med/high \\
 & \texttt{ROUTE\_LAYER\_EFFORT}    & Cat & low/med/high \\
\bottomrule
\end{tabular}
\end{table}

All methods share the same execution stack, quick-to-full protocol, and runtime accounting; only candidate proposal strategies differ.
LLM calls are made only when the candidate buffer is depleted (typically every 4--8 iterations), not at every iteration. Sensitivity and failure analyses are pre-computed into compact text before prompt injection.
Consequently, token growth is tied to buffer-refill points rather than every evaluated trajectory.
Relative to CROP-style retrieval-guided prompting, the main difference is where experience is kept: StateTune accumulates validated summaries in persistent memory and refreshes prompt context only at buffer-refill points, instead of depending as heavily on reassembled online context throughout the search.

Since neither RankTuner~\cite{xu2024ranktuner} nor CROP~\cite{pan2025crop} releases source code targeting our flow, we re-implemented both within our evaluation framework---RankTuner retaining its pairwise GP and Duel-Thompson sampling, CROP using the same RAG-EDA~\cite{pu2024rageda} retrieval as StateTune---so that all methods share identical execution, promotion, and budget-accounting pipelines.

\textbf{Quick-stage endpoint.}
The quick-stage cutoff is a configurable parameter.
We use CTS as the quick-stage endpoint rather than placement alone, because CTS provides richer timing signals that improve the fidelity of quick-stage QoR estimates and enable more effective promotion decisions.

\textbf{Evaluation metrics.}
Best WNS, area, and power are the per-objective extremes observed during the run.
\emph{Final hypervolume} (HV) is the dominated hypervolume of the Pareto frontier at run termination, computed with respect to a fixed reference point \(\mathbf{r}\) that is dominated by all feasible evaluations; it serves as a comprehensive frontier metric~\cite{geng2023ptpt}.
The reference point is determined independently for each comparison group (main comparison, ablation, and sensitivity studies), so absolute HV values are not directly comparable across tables.
\emph{Global Pareto hits} (GP Hits) counts the number of a method's full-valid evaluations that are non-dominated with respect to the union of all methods' evaluations on the same benchmark.
In the main comparison, we focus the narrative on WNS, area, power, and HV; GP hits are used as auxiliary evidence, while time to first global Pareto hit (First Hit) is retained in the ablation analysis.

\section{Results}

This section reports the unified comparison across the benchmark matrix, the ablation study, and dedicated analyses of evidence-gating sensitivity and memory poisoning.

\subsection{Comparison Across the Working Benchmark Matrix}
Table~\ref{tab:results} reports the six benchmark blocks in the full matrix.
We focus on aggregate trends across the full set rather than on any single case.

\begin{table*}[t]
\centering
\caption{Results across the \(2{\times}3\) benchmark matrix under the shared budget caps.
Six methods are compared: BO\,(qEHVI), Optuna-TPE, Random, RankTuner\,(ICCAD'24)~\cite{xu2024ranktuner}, CROP\,(ICCAD'25)~\cite{pan2025crop}, and StateTune.
Bold = best per benchmark (per column).}
\label{tab:results}
\scriptsize
\renewcommand{\arraystretch}{1.06}
\setlength{\tabcolsep}{2.6pt}

\begin{tabular*}{\textwidth}{@{\extracolsep{\fill}}llrrrrr@{\hspace{5mm}}llrrrrr@{}}
\toprule
\multicolumn{7}{c}{ASAP Benchmarks} & \multicolumn{7}{c}{NAN Benchmarks} \\
\cmidrule(r){1-7}\cmidrule(l){8-14}
Design & Method & WNS & Area & Power & GP Hits & HV
& Design & Method & WNS & Area & Power & GP Hits & HV \\
\midrule

\multirow{6}{*}{ASAP-JPEG}
& Random       & -26.1420 & 6042 & 29.9512 & 0  & 1.806e5
& \multirow{6}{*}{NAN-JPEG}
& Random       & -0.0080 & 106638 & 111.9460 & 0  & 1.911e6 \\
& Optuna-TPE   & -13.3730 & 5973 & 29.9231 & 2  & 3.412e5
& & Optuna-TPE   & 0.0130 & 106204 & 111.3026 & 1  & 2.082e6 \\
& BO\,(qEHVI)  & -26.1130 & 6025 & 29.7187 & 1  & 2.348e5
& & BO\,(qEHVI)  & -0.0050 & \textbf{106163} & 111.6088 & 0  & 2.064e6 \\
& RankTuner    & -29.5560 & 6256 & 33.7636 & 0  & 9.938e4
& & RankTuner    & 0.0090 & 108508 & 126.8801 & 0  & 7.849e5 \\
& CROP         & -10.8240 & 6924 & 33.5890 & 1  & 1.286e5
& & CROP         & 0.0120 & 106224 & 111.6372 & 0  & 2.030e6 \\
& \textbf{StateTune} & \textbf{-6.7520} & \textbf{5946} & \textbf{29.4496} & \textbf{12} & \textbf{4.564e5}
& & \textbf{StateTune} & \textbf{0.0150} & \textbf{106163} & \textbf{111.2779} & \textbf{10} & \textbf{2.097e6} \\
\midrule

\multirow{6}{*}{ASAP-AES}
& Random       & -170.7360 & 1678 & 7.9457 & 0  & 3.178e6
& \multirow{6}{*}{NAN-AES}
& Random       & -0.0230 & 22371 & 21.6310 & 1 & 3.887e3 \\
& Optuna-TPE   & -174.7420 & 1670 & 7.7038 & 5  & 3.436e6
& & Optuna-TPE   & -0.0190 & 22379 & 21.5630 & 1 & 4.072e3 \\
& BO\,(qEHVI)  & -201.8410 & 1670 & 7.7462 & 0  & 3.284e6
& & BO\,(qEHVI)  & -0.0180 & 22355 & 21.5925 & 4 & 4.120e3 \\
& RankTuner    & -175.3810 & 1672 & 7.9520 & 0  & 3.233e6
& & RankTuner    & -0.0650 & 22648 & 21.8425 & 0 & 2.357e3 \\
& CROP         & -164.6620 & 1702 & 8.6772 & 0  & 2.633e6
& & CROP         & -0.0250 & 22433 & 21.6294 & 0 & 3.813e3 \\
& \textbf{StateTune} & \textbf{-162.8580} & \textbf{1669} & \textbf{7.5967} & \textbf{10} & \textbf{3.548e6}
& & \textbf{StateTune} & \textbf{-0.0120} & \textbf{22353} & \textbf{21.4920} & \textbf{9} & \textbf{4.316e3} \\
\midrule

\multirow{6}{*}{ASAP-IBEX}
& Random       & -353.5490 & 2063 & 6.5417 & 0  & 2.747e5
& \multirow{6}{*}{NAN-IBEX}
& Random       & 0.0240 & 29030 & 11.6956 & 0  & 1.288e3 \\
& Optuna-TPE   & -258.6870 & 2052 & 6.2220 & \textbf{6} & 4.486e5
& & Optuna-TPE   & 0.0370 & 28943 & 11.6197 & 0  & 1.515e3 \\
& BO\,(qEHVI)  & -335.1390 & 2181 & 7.1288 & 0  & 1.061e5
& & BO\,(qEHVI)  & 0.0590 & \textbf{28928} & 11.5741 & \textbf{9} & 1.780e3 \\
& RankTuner    & -274.0760 & 2061 & 6.2730 & 0  & 3.947e5
& & RankTuner    & 0.0130 & 29068 & 11.7663 & 0  & 1.147e3 \\
& CROP         & -270.8550 & 2063 & 6.5617 & 0  & 3.239e5
& & CROP         & 0.0120 & 29377 & 11.5660 & 0  & 1.113e3 \\
& \textbf{StateTune} & \textbf{-246.9210} & \textbf{2050} & \textbf{6.2107} & 6 & \textbf{4.539e5}
& & \textbf{StateTune} & \textbf{0.0600} & \textbf{28928} & \textbf{11.5597} & 5 & \textbf{1.810e3} \\
\bottomrule
\end{tabular*}
\end{table*}

\begin{figure*}[t]
    \centering
    \includegraphics[width=0.98\textwidth]{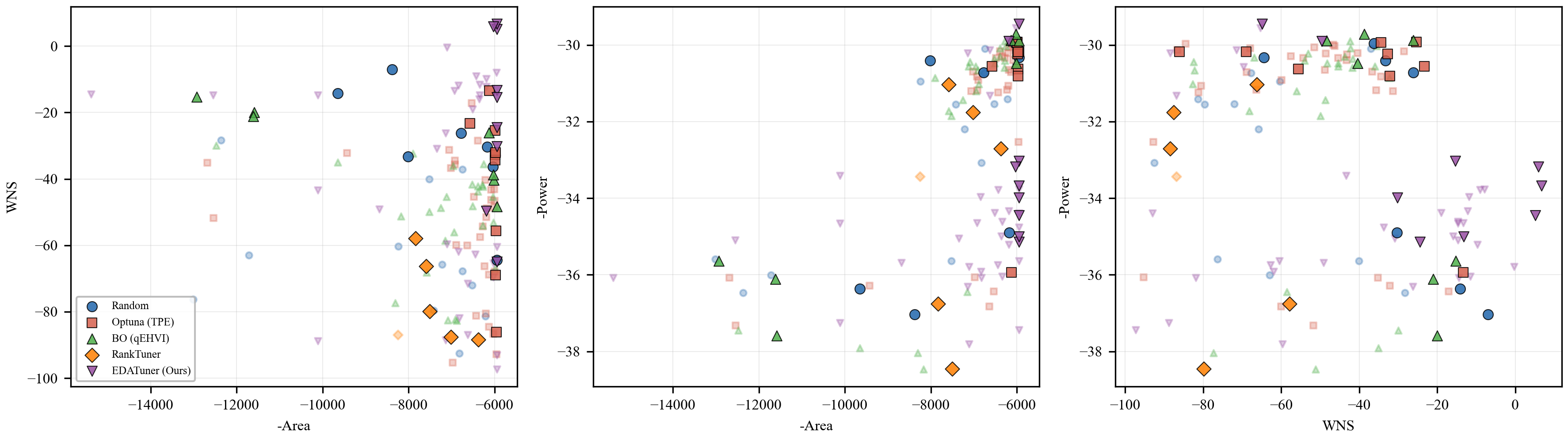}
    \caption{Pareto frontier projections for the ASAP-JPEG case (\(-\)Area vs.\ WNS, \(-\)Area vs.\ \(-\)Power, WNS vs.\ \(-\)Power). StateTune (purple triangles) occupies the strongest region among the compared methods.}
    \label{fig:pareto_proj}
\end{figure*}

\begin{figure}[t]
    \centering
    \includegraphics[width=0.48\textwidth]{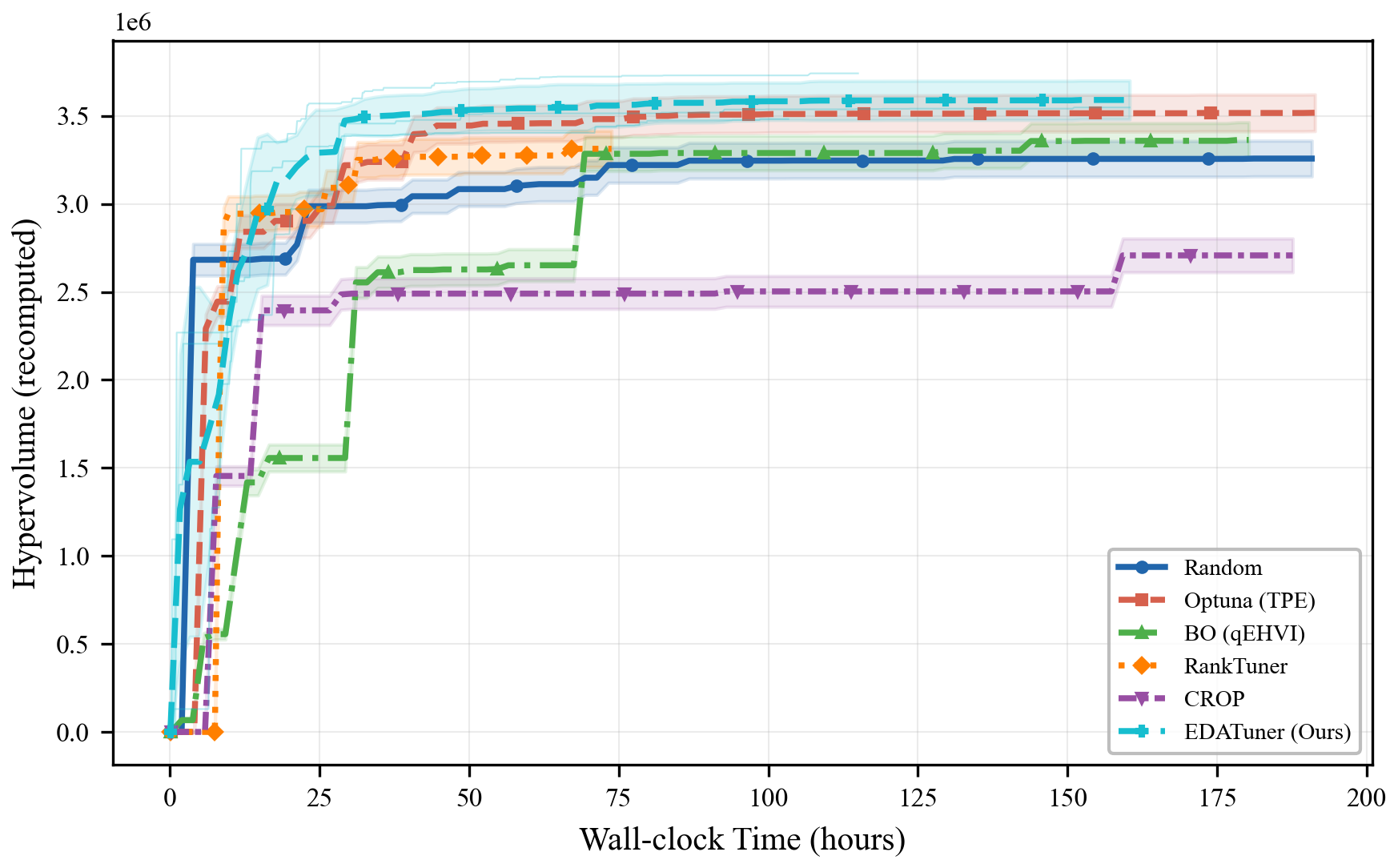}
    \caption{Anytime hypervolume trajectory for ASAP7-AES, averaged over three independent runs per method with shaded \(\pm 1\) standard-deviation bands. StateTune (cyan) separates from all baselines early and maintains a clear advantage throughout the run.}
    \label{fig:hv_traj}
\end{figure}

\textbf{Best per-objective results and frontier quality.}
StateTune attains the best WNS, best power, and best final HV on all six benchmark blocks.
For area, it is the outright best method on ASAP-JPEG, ASAP-IBEX, and NAN-AES, and ties the best value on ASAP-AES, NAN-JPEG, and NAN-IBEX.
Most importantly, the final-HV advantage is consistent across the full benchmark matrix: \(+33.7\%\) over Optuna-TPE on ASAP-JPEG, \(+3.3\%\) over Optuna-TPE on ASAP-AES, \(+1.2\%\) over Optuna-TPE on ASAP-IBEX, \(+0.7\%\) over Optuna-TPE on NAN-JPEG, \(+1.3\%\) over BO\,(qEHVI) on NAN-AES, and \(+1.7\%\) over BO\,(qEHVI) on NAN-IBEX.
Thus, StateTune consistently produces a stronger final Pareto frontier across the full matrix.

\textbf{Auxiliary evidence from GP hits.}
StateTune also leads GP hits on ASAP-JPEG, ASAP-AES, NAN-JPEG, and NAN-AES, and ties the best GP-hit count on ASAP-IBEX.
The GP-hit advantage on these five blocks can be traced to the persistent memory mechanism: accumulated failure patterns and sensitivity signals steer the search away from known-poor regions, concentrating evaluation budget near the Pareto frontier rather than re-exploring dominated configurations.
This separation is useful: on NAN-IBEX, BO attains more GP hits, yet StateTune still delivers the best final HV and the best power, showing that frontier quality is not determined by hit count alone.

\textbf{LLM baselines do not automatically win.}
CROP---despite using an LLM with RAG---remains behind StateTune on all six blocks, and even trails random search on ASAP-JPEG and NAN-IBEX in final HV.
Adding an LLM and retrieval to a tuning loop is not by itself sufficient.
Without evidence gating, noisy LLM suggestions enter the search context unfiltered, progressively biasing the proposal distribution; without a surrogate-guided promotion policy, the system cannot distinguish high-frontier-gain candidates from plausible-looking but dominated ones, wasting the scarce full-evaluation budget.
RankTuner\,(ICCAD'24)~\cite{xu2024ranktuner} also trails on every block: its preference-based ranking struggles in this mixed continuous-categorical space with 19 parameters, the pairwise GP scales poorly with dimensionality, and without cross-stage knowledge transfer or persistent memory, each run starts from scratch.

\textbf{Reproducibility.}
To assess run-to-run stability, we repeat StateTune three times on every benchmark block under identical budgets.
Table~\ref{tab:repro} reports the HV mean and standard deviation; Figure~\ref{fig:hv_traj} visualizes the anytime trajectory for ASAP7-AES.
The coefficient of variation (CV) remains below 7\% on five of six blocks (e.g., 0.89\% on ASAP7-AES and 0.51\% on NAN-JPEG), confirming reproducibility across seeds.
ASAP7-JPEG exhibits the highest CV (15.4\%), but even its worst single-seed HV (\(\approx 4.5 \times 10^5\)) still exceeds all baselines in Table~\ref{tab:results}.
In all six cases, the repeated-run mean matches or exceeds the single-run HV in Table~\ref{tab:results}, indicating that the main comparison is not an outlier.

\begin{table}[t]
\centering
\caption{StateTune reproducibility over three independent runs.}
\label{tab:repro}
\footnotesize
\setlength{\tabcolsep}{3.5pt}
\renewcommand{\arraystretch}{1.10}

\begin{tabular}{lcc@{\hspace{10pt}}lcc}
\toprule
\multicolumn{3}{c}{\textbf{ASAP7}} & \multicolumn{3}{c}{\textbf{NAN}} \\
\cmidrule(r){1-3}\cmidrule(l){4-6}
Benchmark & HV (\(\mu \pm \sigma\)) & CV (\%)
& Benchmark & HV (\(\mu \pm \sigma\)) & CV (\%) \\
\midrule
JPEG & \(5.33e5 \pm 8.21e4\) & 15.4
& JPEG & \(2.53e6 \pm 1.29e4\) & 0.51 \\
AES  & \(3.73e6 \pm 3.32e4\) & 0.89
& AES  & \(4.42e3 \pm 6.34e1\) & 1.43 \\
IBEX & \(4.57e5 \pm 2.86e4\) & 6.27
& IBEX & \(1.66e3 \pm 9.00e1\) & 5.42 \\
\bottomrule
\end{tabular}
\end{table}

\subsection{Ablation Study}
\label{sec:ablation}

\begin{table}[t]
\centering
\caption{Ablation on ASAP7 JPEG under 48\,thread\(\cdot\)h budget. Variants ordered by degradation severity.}
\label{tab:ablation}
\scriptsize
\setlength{\tabcolsep}{4.0pt}
\begin{tabular}{lcccc}
\toprule
Variant & Final HV (M) & HV Drop (\%) & Pareto Hits & First Hit (h) \\
\midrule
Full system      & 10.10 & ---   & 6 & 10.29 \\
w/o Persistence  & 4.19  & 58.5  & 0 & -- \\
w/o KA           & 4.57  & 54.8  & 2 & 2.10 \\
w/o RAG          & 5.44  & 46.1  & 0 & -- \\
w/o CoT          & 6.92  & 31.5  & 1 & 29.62 \\
w/o EHVI         & 8.32  & 17.6  & 4 & 12.93 \\
\bottomrule
\end{tabular}
\end{table}

Each ablation variant removes one component while keeping the rest intact (Table~\ref{tab:ablation}).

\textbf{w/o Persistence.}
Persistent memory propagates accumulated failure patterns, constraints, and sensitivity signals across iterations.
Without it, HV drops to 4.19\,M (\(-58.5\%\)) with zero Pareto hits---despite retaining LLM, RAG, CoT, and EHVI.
This confirms that persistent memory is the largest single contributor and a distinct design axis from LLM-based search.

\textbf{w/o KA.}
The knowledge agent translates memory into context-aware proposals conditioned on failures, sensitivities, and the design prior.
Without it, HV falls to 4.57\,M (\(-54.8\%\)) and only two Pareto hits remain.
Together with the persistence ablation, this confirms that both memory and LLM guidance are necessary---neither alone suffices.

\textbf{w/o RAG.}
RAG grounds proposals in EDA tool documentation and design-specific best practices; without it, CoT, memory, and EHVI remain intact but the LLM relies solely on parametric knowledge.
HV drops to 5.44\,M (\(-46.1\%\)) with zero Pareto hits, confirming that general LLM knowledge is insufficient without retrieval-augmented domain grounding.

\textbf{w/o CoT.}
CoT reasoning structures failure diagnosis and sensitivity assessment into analytical traces that inform proposals.
Without it, HV drops to 6.92\,M (\(-31.5\%\)), confirming that RAG supplies domain context but CoT supplies the reasoning machinery to act on it.

\textbf{w/o EHVI.}
EHVI-guided promotion ranks candidates by expected frontier gain per unit of runtime cost.
Without it, the first global Pareto hit is delayed from 10.29\,h to 12.93\,h and HV falls to 8.32\,M (\(-17.6\%\)), confirming that unguided promotion wastes the scarce full-evaluation budget on dominated proposals.

Overall, persistent memory is the primary contributor, while EHVI-guided promotion provides a complementary gain.

\subsection{Sensitivity to Evidence-Gating Threshold \(k\)}
\label{sec:ksensitivity}

\begin{figure}[t]
    \centering
    \begin{subfigure}[t]{0.48\columnwidth}
        \centering
        \includegraphics[width=\linewidth]{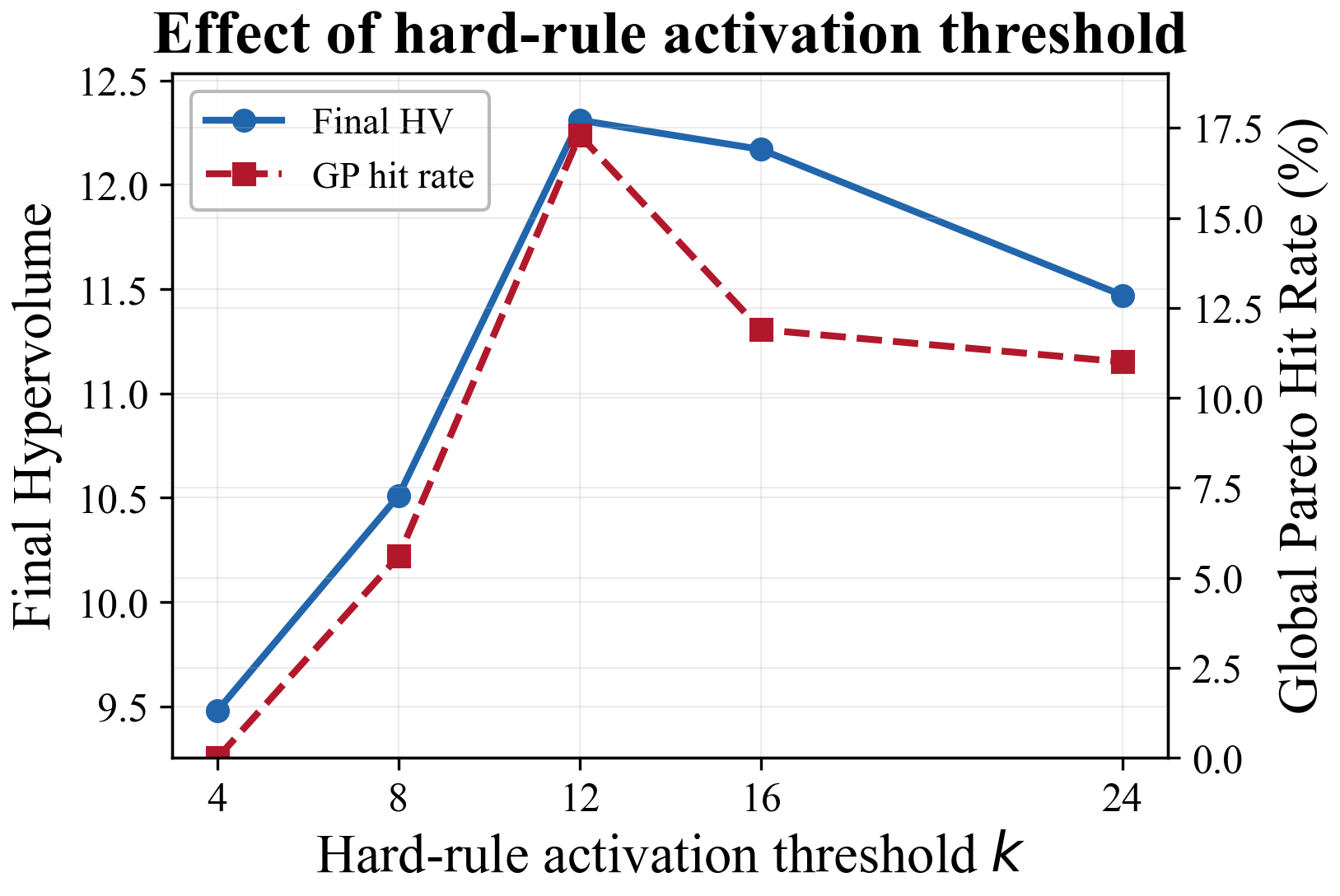}
        \caption{Evidence-gating sensitivity.}
        \label{fig:k_vs_hv}
    \end{subfigure}\hfill
    \begin{subfigure}[t]{0.48\columnwidth}
        \centering
        \includegraphics[width=\linewidth]{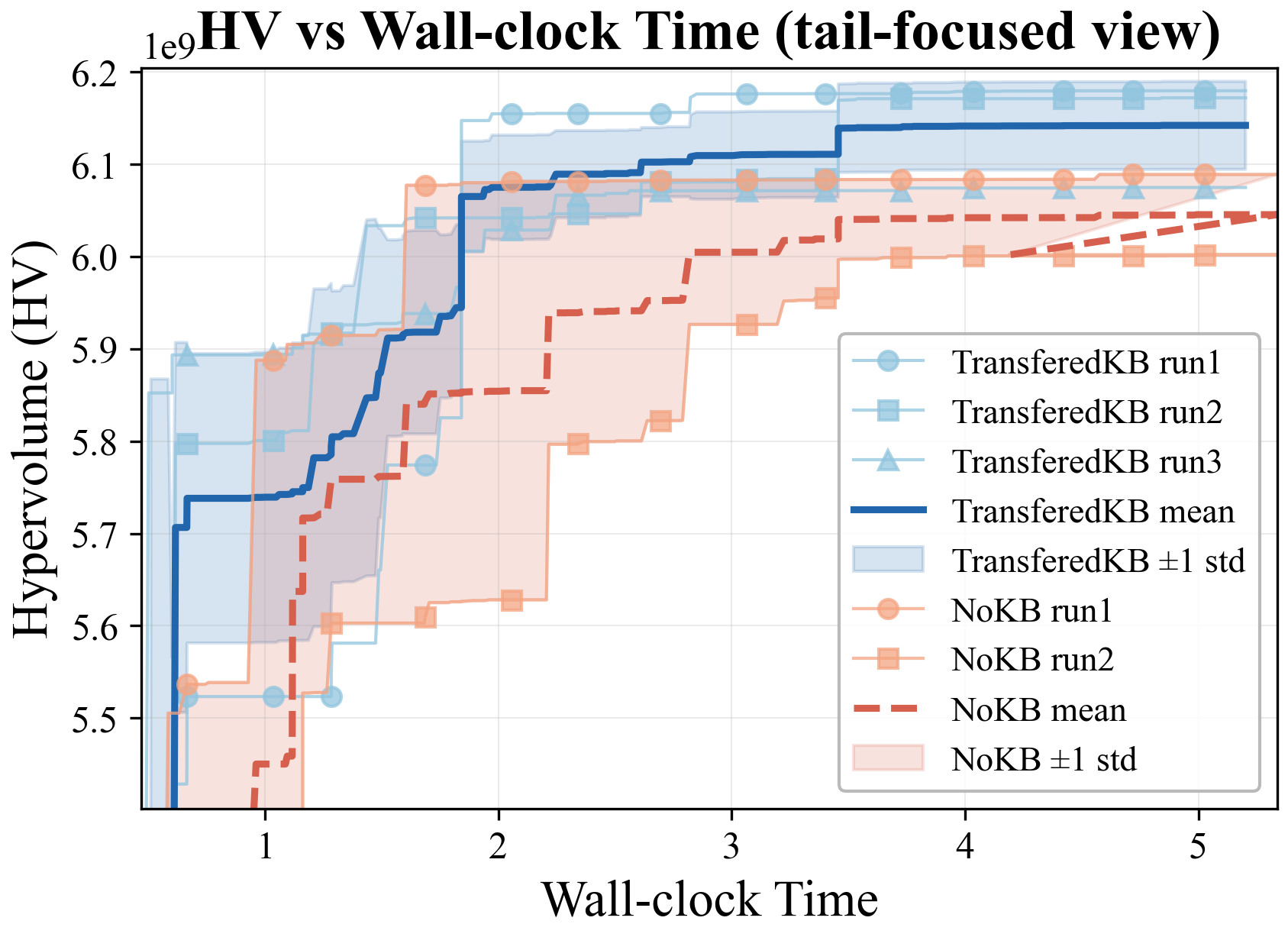}
        \caption{Cross-design transfer.}
        \label{fig:transfer}
    \end{subfigure}

    \captionsetup{belowskip=-8pt}
    \caption{Additional validation on ASAP7 JPEG under 48\,thread\(\cdot\)h. (a) Final HV and global Pareto hit rate peak at \(k{=}12\). (b) Transferred memory converges faster and reaches higher final HV than no transfer. Thin curves denote individual runs, thick curves denote means, and shaded bands indicate \(\pm 1\) standard deviation.}
    \label{fig:k_ablation}
\end{figure}

The evidence-gating threshold \(k\) controls how many corroborating full-valid observations are required before a hard rule becomes active.
To experimentally validate the choice of \(k{=}12\), we sweep \(k \in \{4, 8, 12, 16, 24\}\) on ASAP7 JPEG under the 48\,thread\(\cdot\)h ablation budget (Figure~\ref{fig:k_vs_hv}).

HV rises steeply from 9.48\,M at \(k{=}4\) to a peak of 12.31\,M at \(k{=}12\) (Pareto hit rate 17.3\%), then declines to 11.47\,M at \(k{=}24\).
This non-monotonic pattern confirms that \(k{=}12\) balances premature activation of incorrect rules against delayed activation of useful rules.

\subsection{Memory Poisoning}
\label{sec:poisoning}

To test whether ungated memory accumulates errors, we compare \emph{Pure} (default evidence gating) and \emph{Poisoned} (all LLM-emitted rules admitted without validation) on ASAP7 JPEG under a 48\,thread\(\cdot\)h budget.

The Poisoned variant suffers substantial degradation: full-valid rate drops from 5.42\% to 2.31\%, and the validated-rule rate falls from 30.4\% to 17.4\% while the contradicted-rule rate rises from 0\% to 17.4\%.
These results confirm that persistent memory without evidence gating accumulates incorrect rules and materially harms search quality, validating the gating mechanism as a necessary component rather than an optional filter.

\subsection{Cross-Design Knowledge Transfer}
\label{sec:transfer}

Because the persistent memory stores typed, design-agnostic artifacts (e.g., parameter constraints, sensitivity patterns, and failure modes), it can be transferred across designs.
We test this by aggregating validated memory from the four non-JPEG benchmark blocks, ranking artifacts by a transfer-confidence score, and using the retained top-\(k\) artifacts to initialize a new ASAP7-JPEG run under the same 48\,thread\(\cdot\)h budget.

The effect is consistent across both seeds (Figure~\ref{fig:transfer}).
Transferred runs enter the \(6.10\text{--}6.12\times10^{9}\) HV band and remain there, whereas no-transfer runs plateau lower at about \(5.95\times10^{9}\) and \(6.03\times10^{9}\).
The transferred mean also reaches \(6.0\times10^{9}\) earlier, at around \(1.8\text{--}1.9\) hours, while the no-transfer mean stays below that level until near termination.
By the end of the run, transfer improves mean HV from about \(5.99\times10^{9}\) to \(6.12\times10^{9}\), an absolute gain of roughly \(1.3\times10^{8}\).
Overall, transferred memory improves both convergence speed and final-HV stability across repeated runs.

\subsection{Limitations}

The six-block matrix spans two technology nodes and JPEG/AES/IBEX, showing consistent cross-case gains; broader benchmarks and industrial SoC studies would further strengthen external validity.
The flat memory structure could be extended to hierarchical or graph-based organizations~\cite{xu2025amem,kang2025memoryos}, which may help when memory grows to thousands of rules across many designs.

\section{Conclusion}

StateTune formulates LLM-assisted EDA flow tuning as a closed-loop, state-carrying process in which a typed, evidence-gated persistent memory jointly supports candidate generation and EHVI-guided promotion.
Across six benchmark blocks and five baselines, it achieves the best final hypervolume on every block.
Ablation shows that persistent memory is the primary contributor (\(-58.5\%\) HV when removed), while EHVI promotion provides a secondary gain (\(-17.6\%\)).
Evidence gating, cross-design transfer, and reproducibility experiments further validate the proposed memory design. Future work includes evaluation on larger industrial designs and richer memory organizations for scaling across more benchmark families.

\begin{acks}
This work is supported by National Science and Technology Major Project (2021ZD0114701).
\end{acks}

\bibliographystyle{ACM-Reference-Format}
\bibliography{reference}

\end{document}